\newif\ifarxiv
\arxivtrue

\documentclass[conference]{IEEEtran}
\IEEEoverridecommandlockouts
\usepackage{cite}
\usepackage{amsmath,amssymb,amsfonts}
\usepackage{algorithmic}
\usepackage{graphicx}
\usepackage{textcomp}
\usepackage{multirow}
\usepackage{xcolor}
\usepackage{booktabs} 
\usepackage{hyperref} 
\def\BibTeX{{\rm B\kern-.05em{\sc i\kern-.025em b}\kern-.08em
    T\kern-.1667em\lower.7ex\hbox{E}\kern-.125emX}}
\begin{document}

\title{QKAN-LSTM: Quantum-inspired Kolmogorov-Arnold Long Short-term Memory

\ifarxiv
\thanks{
The views expressed in this article are those of the authors and do not represent the views of Wells Fargo. This article is for informational purposes only. Nothing contained in this article should be construed as investment advice. Wells Fargo makes no express or implied warranties and expressly disclaims all legal, tax, and accounting implications related to this article.\\
\IEEEauthorrefmark{1} \href{mailto:ycchen1989@ieee.org}{ycchen1989@ieee.org}.
\IEEEauthorrefmark{2} \href{mailto:kuoenjui@nycu.edu.tw}{kuoenjui@nycu.edu.tw}.
\IEEEauthorrefmark{3} \href{mailto:goan@phys.ntu.edu.tw}{goan@phys.ntu.edu.tw}.
}
\fi
}

\ifarxiv
\author{
\IEEEauthorblockN{
    Yu-Chao Hsu\IEEEauthorrefmark{4}\IEEEauthorrefmark{5},
    Jiun-Cheng Jiang\IEEEauthorrefmark{4}\IEEEauthorrefmark{6}\IEEEauthorrefmark{7},
    Chun-Hua Lin\IEEEauthorrefmark{4}\IEEEauthorrefmark{6},
    Kuo-Chung Peng\IEEEauthorrefmark{4}\IEEEauthorrefmark{6},\\
    Nan-Yow Chen\IEEEauthorrefmark{4},
    Samuel Yen-Chi Chen\IEEEauthorrefmark{8}\IEEEauthorrefmark{1}, 
    En-Jui Kuo\IEEEauthorrefmark{9}\IEEEauthorrefmark{2},
    Hsi-Sheng Goan\IEEEauthorrefmark{6}\IEEEauthorrefmark{7}\IEEEauthorrefmark{10}\IEEEauthorrefmark{3}
}
\IEEEauthorblockA{\IEEEauthorrefmark{4} National Center for High-Performance Computing, National Institutes of Applied Research, Hsinchu, Taiwan}
\IEEEauthorblockA{\IEEEauthorrefmark{5} Cross College Elite Program, National Cheng Kung University, Tainan, Taiwan}
\IEEEauthorblockA{\IEEEauthorrefmark{6} Department of Physics and Center for Theoretical Physics, National Taiwan University, Taipei, Taiwan}
\IEEEauthorblockA{\IEEEauthorrefmark{7} Center for Quantum Science and Engineering, National Taiwan University, Taipei, Taiwan}
\IEEEauthorblockA{\IEEEauthorrefmark{8}Wells Fargo, New York, NY, USA}
\IEEEauthorblockA{\IEEEauthorrefmark{9}Department of Electrophysics, National Yang Ming Chiao Tung University, Hsinchu, Taiwan}
\IEEEauthorblockA{\IEEEauthorrefmark{10}Physics Division, National Center for Theoretical Sciences, Taipei, Taiwan}
}
\else
\author{
Anonymous Authors
}
\fi

\maketitle

\begin{abstract}
Long short-term memory (LSTM) models are a particular type of recurrent neural networks (RNNs) that are central to sequential modeling tasks in domains such as urban telecommunication forecasting, where temporal correlations and nonlinear dependencies dominate.
However, conventional LSTMs suffer from high parameter redundancy and limited nonlinear expressivity.
In this work, we propose the Quantum-inspired Kolmogorov–Arnold Long Short-Term Memory (QKAN-LSTM), which integrates Data Re-Uploading Activation (DARUAN) modules into the gating structure of LSTMs.
Each DARUAN acts as a quantum variational activation function (QVAF), enhancing frequency adaptability and enabling an exponentially enriched spectral representation without multi-qubit entanglement.
The resulting architecture preserves quantum-level expressivity while remaining fully executable on classical hardware.
Empirical evaluations on three datasets, Damped Simple Harmonic Motion, Bessel Function, and Urban Telecommunication, demonstrate that QKAN-LSTM achieves superior predictive accuracy and generalization with a 79\% reduction in trainable parameters compared to classical LSTMs.
We extend the framework to the Jiang–Huang–Chen–Goan Network (JHCG Net), which generalizes KAN to encoder–decoder structures, and then further use QKAN to realize the latent KAN, thereby creating a Hybrid QKAN (HQKAN) for hierarchical representation learning.
The proposed HQKAN-LSTM thus provides a scalable and interpretable pathway toward quantum-inspired sequential modeling in real-world data environments.
\end{abstract}


\section{Introduction}

Machine learning (ML) has achieved remarkable success across diverse domains, with recurrent neural networks (RNNs) and their gated variants, such as the long short-term memory (LSTM) network~\cite{6795963}, forming the backbone of sequence modeling and temporal prediction tasks~\cite{6707742,sutskever2014sequencesequencelearningneural,PhysRevA.95.012335,Flurin_2020}. 
Among these, LSTMs have demonstrated exceptional ability to capture nonlinear temporal dynamics and long-range dependencies, making them indispensable tools in modeling complex spatiotemporal systems, including those in telecommunication networks~\cite{greff2017lstm,LINDEMANN2021650,chen2025benchmarking}. 
In urban telecommunication systems, LSTMs are particularly valuable for forecasting user activity patterns and network loads from historical time series, where data often exhibit irregular periodicity, bursty behavior, and strong spatial-temporal correlations~\cite{chen2025benchmarking}. 
Accurate telecommunication forecasting is crucial for real-time network resource allocation, traffic optimization, and anomaly detection in large-scale urban environments~\cite{barlacchi2015multi}.

Despite their success, conventional LSTMs face inherent challenges related to vanishing gradients, high computational overhead, and overparameterization, which limit scalability and interpretability when applied to high-frequency, high-dimensional telecommunication data. 
Moreover, the reliance on static activation functions constrains their representational richness, particularly when modeling complex oscillatory patterns and nonlinear feedback prevalent in communication signals.

In parallel, quantum machine learning (QML) has emerged as a promising paradigm that utilizes the principles of quantum mechanics, such as superposition, interference, and entanglement, to enhance functional expressivity and parameter efficiency~\cite{chen2020variational,phillipson2020quantum,Schuld_2021,chen2022quantum,watkins2023quantum,qu2024qmfnd,chen2025jet,liu2025qttf}.
Some QML methods, such as the quantum kernel method~\cite{chen2024support,tsai2025learning,hsu2025kernel,hsu2025climate,liu2025relational,chen2025validating}, have also been developed and explored. 
However, current QML implementations remain limited by noisy intermediate-scale quantum (NISQ) hardware, constrained qubit counts, and insufficient two-qubit gate fidelities~\cite{liu2024training,Singh_2024,PRXQuantum.5.040342,liu2025quantum,42w2-6ccy}, restricting their scalability in real-world applications such as telecommunication signal prediction.

To bridge the gap between classical and quantum paradigms, recent efforts have focused on \emph{quantum-inspired architectures} that retain the expressive power of quantum models while remaining executable on classical hardware. 
The quantum-inspired Kolmogorov–Arnold network (QKAN)~\cite{jiang2025qkan} exemplifies this approach. 
QKAN reinterprets the Kolmogorov–Arnold network (KAN)~\cite{liu2025kan} by employing single-qubit data re-uploading circuits~\cite{Mitarai_2018,Schuld_2021} as \emph{quantum variational activation functions} (QVAFs), effectively forming a \emph{DatA Re-Uploading ActivatioN} (DARUAN). 
Each DARUAN module encodes input features into parameterized rotations on a single-qubit Bloch sphere, with trainable pre-processing weights in each data-uploading block, enabling an exponentially rich Fourier representation without the need for multi-qubit entanglement.
The method of adding trainable pre-processing weights has also been employed for multi-qubits data re-uploading circuits~\cite{yu2022powerlimitationssinglequbitnative,zhao2024quantum}.
This design allows QKANs to preserve quantum-level expressivity while remaining computationally tractable from single CPU to multi-nodes high-performance computing (HPC) GPU clusters.
While QKANs remain feasible for current real quantum devices, state-of-the-art quantum devices have already empirically achieved single-qubit gate error rates at the $10^{-5}$--$10^{-7}$ scale, including superconducting devices~\cite{PRXQuantum.5.040342}, spin qubit devices~\cite{wu2025simultaneoushighfidelitysinglequbitgates}, and trapped-ion devices~\cite{42w2-6ccy}.

Integrating QKANs into recurrent architectures such as LSTMs introduces a powerful new class of hybrid models—\textbf{QKAN-LSTMs}—that unify the temporal modeling strength of LSTMs with the spectral expressivity of quantum-inspired activations. 
In this framework, QKANs replace the classical feedforward layers within the LSTM cell, acting as adaptive, quantum-enhanced feature extractors and parameter compressors. 
Recent studies have demonstrated that such hybrid quantum-classical architecture improves sequence modeling efficiency and generalization across time-series forecasting and natural language generation tasks~\cite{chen2020quantumlongshorttermmemory,hsu2025federated,jiang2025qkan,chen2025quantumlongshorttermmemory}. 
By leveraging the compact harmonic representation of DARUAN-based QKANs, QKAN-LSTMs achieve enhanced trainability, reduced parameter counts, and robustness against gradient degradation—offering a scalable and physically interpretable pathway toward efficient sequential modeling.

Beyond temporal sequence modeling, the KAN framework has also been generalized to hierarchical architectures through the Jiang–Huang–Chen–Goan Network (JHCG Net)~\cite{jiang2025qkan}. 
The JHCG Net extends the KAN paradigm into an encoder–KAN–decoder topology, where the latent KAN module serves as a nonlinear feature processor within an autoencoder-like structure. 
When the latent processor is realized using QKANs, the resulting \textit{hybrid quantum-inspired Kolmogorov–Arnold network (HQKAN)} architecture integrates quantum-inspired transformations into the latent feature space, enabling exponential Fourier frequency spectral enrichment with less width and depth compared to classical KAN and MLP. 
Crucially, HQKANs function as scalable, drop-in replacements for multilayer perceptrons (MLPs) in deep architectures such as Transformers and Diffusion Models, maintaining classical differentiability and GPU compatibility while offering superior expressivity and efficiency.

In this work, we extend the QKAN framework to sequential modeling by integrating it into the LSTM architecture, forming the QKAN-LSTM. 
We further position this model within the broader hybrid paradigm defined by the HQKAN, which generalizes QKANs into hierarchical encoder–decoder architectures for representation learning.
Together, QKAN-LSTMs and HQKANs establish a unified framework for quantum-inspired learning that spans both temporal modeling and hierarchical feature representation.

\noindent We summarize our main contributions as follows:

\begin{enumerate}
    \item We introduce a novel QKAN-LSTM architecture that integrates quantum-inspired DARUAN modules within LSTM cells, replacing conventional affine transformations to enhance nonlinear expressivity and parameter efficiency. 
    Furthermore, we extend this framework to its hybrid counterpart, HQKAN-LSTM, which embeds the JHCG Net mechanism for additional scalability and compression efficiency.
    \item We achieve a substantial \textbf{79\% reduction in trainable parameters} compared to classical LSTMs while maintaining or improving predictive performance across multiple datasets.
    \item We evaluate the proposed models on three representative benchmarks—damped harmonic motion, Bessel function regression, and urban telecommunication forecasting—demonstrating superior accuracy, stability, and generalization compared to both LSTM and QLSTM baselines.
\end{enumerate}

\section{Related Work}

\paragraph{Quantum-enhanced long short-term memory}
Ref.~\cite{ceschini2021realhd} presents a fully quantum implementation of LSTM cells directly within quantum circuits.
Quantum enhanced LSTM variants integrate LSTM cell with variational circuits~\cite{chen2022quantum}, quantum kernels~\cite{hsu2025kernel,hsu2025climate} or quantum convolutional networks~\cite{xu2024qcnn}, while Quantum-Train LSTM replaces LSTM trainable parameters with quantum circuit outputs~\cite{liu2025qtlstm,lin2024qtlstm}.
Applications are also explored in telecommunication~\cite{chen2025benchmarking}, weather prediction~\cite{liu2025qttf,lin2024qtlstm,hsu2025climate}, cosmology~\cite{liu2025qtlstm}, fraud detection~\cite{ubale2025FD}, traffic~ \cite{saini2024traffic}, solar power~\cite{Khan_2024}, stress monitoring~\cite{padha2022stress} and indoor localization~\cite{chien2024indoor}.

\paragraph{Kolmogorov-Arnold network and its applications for time series forecasting} Liu et al.~\cite{liu2025kan} introduced KANs, a neural architecture inspired by the Kolmogorov-Arnold representation theorem (KART)~\cite{kolmogorov1957representation}.
Refs.~\cite{liu2025kan,liu2024kan20kolmogorovarnoldnetworks} generalized KART to arbitrary widths and depths and showed that KANs are able to replace MLPs with better accuracy and interpretability.
Subsequent studies have applied KANs to time series forecasting tasks~\cite{vacarubio2024kolmogorovarnoldnetworkskanstime,xu2024kolmogorovarnoldnetworkstimeseries,genet2024tkantemporalkolmogorovarnoldnetworks,genet2024temporalkolmogorovarnoldtransformertime,lstm-kan-1,yu2025lstmkan, gong2024research,cui2025real}, confirming their effectiveness in modeling temporal data.

\section{Methodology}

\subsection{Quantum-inspired Kolmogorov-Arnold Long Short-term Memory}
To further enhance the nonlinear modeling capability of LSTM networks, we propose the QKAN-LSTM model.
This architecture replaces the conventional affine transformations in LSTM gates with quantum-inspired functional modules based on the KART.
By constructing each gate as a composition of multiple variational quantum subfunctions acting on individual input dimensions, the QKAN-LSTM approximates complex high-dimensional nonlinear mappings through a structured aggregation of one-dimensional quantum transformations.
This design enables stronger nonlinear expressivity and improved long-range dependency modeling in sequential data.

\begin{figure*}[!t]
    \centering
    \includegraphics[width=\textwidth]{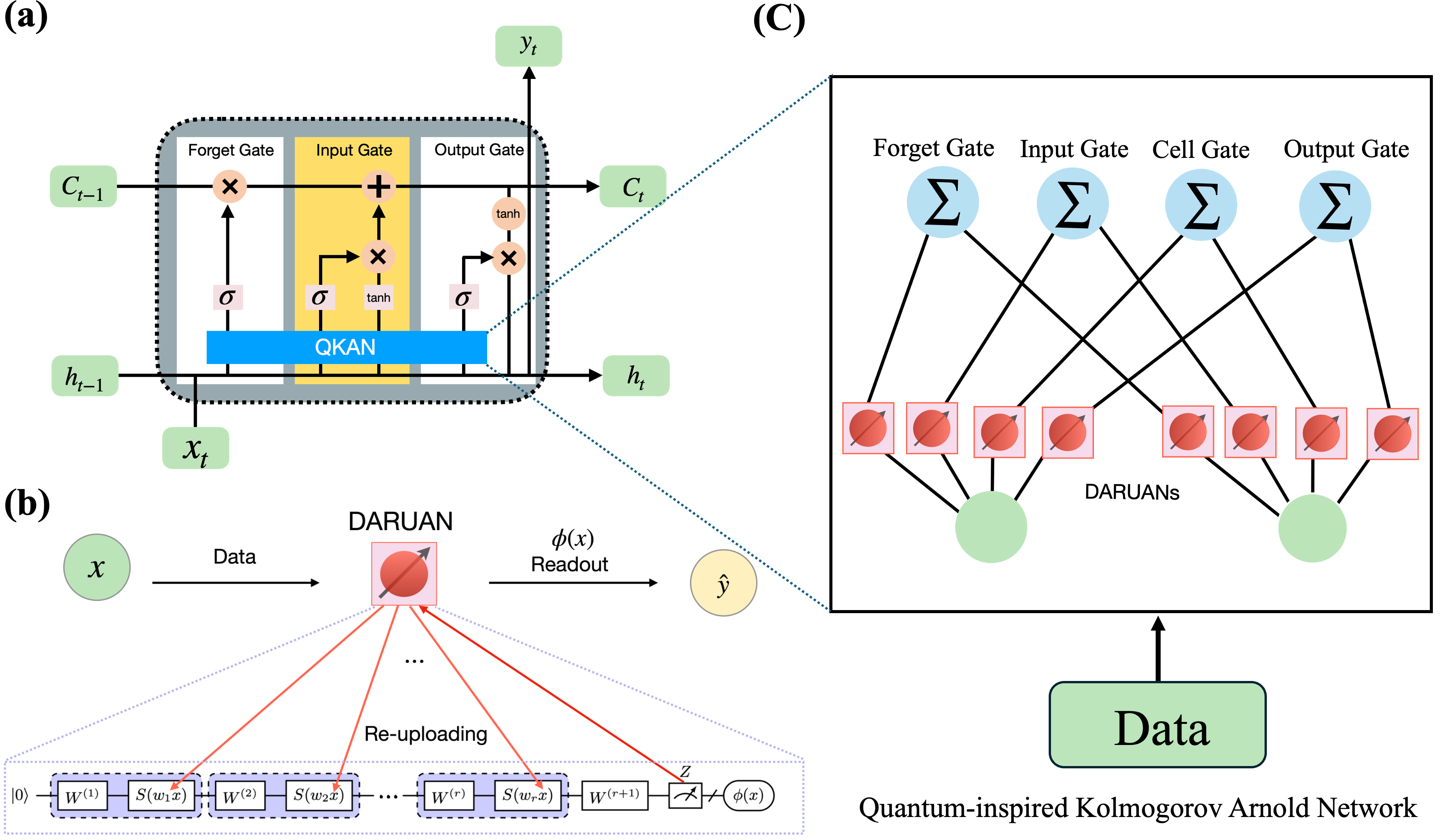}
    \caption{\textbf{Overview of the QKAN-LSTM architecture.}
    \textbf{(a)} The architecture of the QKAN-LSTM model with QKAN integration in the input, forget, cell, and output gates.
    \textbf{(b)} The data is fed into the DARUAN layer, where the quantum features are re-uploaded and processed.
    \textbf{(c)} A detailed view of how the DARUANs are applied to the gates in QKAN to enhance the LSTM's ability to capture complex and non-linear sequence dependencies with QVAF.
    }
    \label{QKAN-LSTM}
\end{figure*}

As illustrated in Figure~\ref{QKAN-LSTM}, this design enriches the expressive capacity of the recurrent dynamics and improves the modeling of long-range temporal dependencies in sequential data.

\subsubsection{Classical LSTM Equations}
The conventional LSTM cell consists of three primary gates—the forget gate (\( f_t \)), input gate (\( i_t \)), and output gate (\( o_t \))—along with a memory cell state (\( C_t \)).  
The evolution of these components over time is governed by the following equations:
\begin{subequations}
\begin{align}
f_t &= \sigma\!\left( W_f [h_{t-1}, x_t] + b_f \right), \label{eq:classical_forget_gate} \\
i_t &= \sigma\!\left( W_i [h_{t-1}, x_t] + b_i \right), \label{eq:classical_input_gate} \\
\tilde{C}_t &= \tanh\!\left( W_C [h_{t-1}, x_t] + b_C \right), \label{eq:classical_candidate_cell_state} \\
C_t &= f_t \odot C_{t-1} + i_t \odot \tilde{C}_t, \label{eq:classical_cell_state_update} \\
o_t &= \sigma\!\left( W_o [h_{t-1}, x_t] + b_o \right), \label{eq:classical_output_gate} \\
h_t &= o_t \odot \tanh(C_t), \label{eq:classical_hidden_state}
\end{align}
\end{subequations}
where:
\begin{itemize}
    \item \( x_t \in \mathbb{R}^n \) represents the $n$-dim input vector at time step \( t \),
    \item \( h_{t-1} \in \mathbb{R}^m \) denotes the $m$-dim hidden state from the previous step,
    \item \( W_f, W_i, W_C, W_o \) are learnable weight matrices associated with each gate,
    \item \( b_f, b_i, b_C, b_o \) are the corresponding bias vectors,
    \item \( \sigma(\cdot) \) and \( \tanh(\cdot) \) are the sigmoid and hyperbolic tangent activation functions, respectively,
    \item \( \odot \) denotes element-wise multiplication.
\end{itemize}

\subsubsection{Integration of Quantum-inspired Kolmogorov--Arnold Networks into LSTM}
In QKAN-LSTM architecture,
the conventional affine transformations \( W[h_{t-1}, x_t] + b \) in the LSTM gates
are replaced by QKAN modules.
Instead of a single linear mapping, each gate aggregates \emph{edge-wise}
quantum variational activation units, following the Kolmogorov--Arnold principle
to approximate high-dimensional nonlinear functions through a sum of
learnable one-dimensional mappings.
This design enables each gate to process inputs through a set of QVAFs, effectively enriching the nonlinear mapping space without altering the classical neuron structure.
Consequently, the QKAN-LSTM exhibits enhanced expressive capacity and spectral diversity in its recurrent dynamics.

Let the concatenated gate input vector be defined as
\[
v_t = [h_{t-1};\, x_t] \in \mathbb{R}^{d}, \qquad d = n + m,
\]
where \([\,\cdot;\cdot\,]\) denotes vector concatenation.
For each gate \( g \in \{f,i,C,o\} \), a QKAN layer mapping is formulated as
\begin{equation}
\label{eq:qkan_gate_map}
\Phi_g(v_t;\Theta_g)
= \sum_{p=1}^{\alpha}
\phi_{g,p}\!\big(v_{t};\,\boldsymbol{\theta}_{g,p}\big),
\end{equation}
where each quantum subnetwork \(\phi_{g,p}(\cdot;\boldsymbol{\theta}_{g,p})\)
serves as a trainable nonlinear activation function realized by \emph{DARUAN}.

The QVAF for each edge unit is defined as
\begin{equation}
\label{eq:qvaf_output}
\phi_{g,p}(u;\boldsymbol{\theta})
= \big\langle 0 \big|
U(u;\boldsymbol{\theta})^{\dagger}\,
M\,
U(u;\boldsymbol{\theta})
\big| 0 \big\rangle ,
\end{equation}
where \( M \) is a fixed Hermitian observable (e.g., \( \sigma_z \)).
The corresponding data re-uploading circuit \( U(u;\boldsymbol{\theta}) \)
consists of a sequence of \(r\) parameterized quantum blocks:
\begin{equation}
\label{eq:qvaf_circuit}
U(u;\boldsymbol{\theta})
= W^{(r+1)}\prod_{\ell=r}^{1}
\Big[
\exp\!\big(-i\,\tfrac{a^{(\ell)}u+b^{(\ell)}}{2}\,G\big)W^{(\ell)}\
\Big].
\end{equation}

Here \(G\) denotes a fixed Hermitian generator,
\(a^{(\ell)}\) and \(b^{(\ell)}\) are scalar encoding parameters,
and \(W^{(\ell)}(\boldsymbol{\theta})\) represents a trainable
single-qubit variational block.
Stacking \(r\) such re-uploading layers with the help of \(a^{(\ell)}\) and \(b^{(\ell)}\) endows the activation
\(\phi_{g,p}\) with an exponentially enriched Fourier spectrum,
enabling compact yet expressive nonlinear representations.

In our work, we initialize the quantum state using a Hadamard gate,
preparing the system in a uniform superposition.
The DURAUN operation is instantiated as
\begin{equation}
S_\ell(w_\ell u)
= R_z(w_\ell u)
= \exp\!\left(-i\,\frac{w_\ell u}{2}\sigma_z\right),
\end{equation}
which corresponds to Eq.~\eqref{eq:qvaf_circuit} with
\(H=\sigma_z\), \(a^{(\ell)} = w_\ell\), and \(b^{(\ell)} = 0\).
The trainable variational block is chosen as
\begin{equation}
W^{(\ell)}(\boldsymbol{\theta})
=\exp\!\left(-i\,\frac{\theta_\ell^{(y)}}{2}\sigma_y \right)
  \exp\!\left(-i\,\frac{\theta_\ell^{(z)}}{2}\sigma_z \right),
\end{equation}
where $\sigma_y$ is the Pauli-Y operator and $\sigma_z$ is the Pauli-Z operator.

Given the QKAN mappings \(\Phi_g(v_t;\Theta_g)\),
the LSTM gating dynamics remain structurally identical to the classical case:
\begin{subequations}
\label{eq:qkan_lstm}
\begin{align}
f_t &= \sigma\!\big(\Phi_f(v_t;\Theta_f)\big), \\
i_t &= \sigma\!\big(\Phi_i(v_t;\Theta_i)\big), \\
\tilde{C}_t &= \tanh\!\big(\Phi_C(v_t;\Theta_C)\big), \\
C_t &= f_t \odot C_{t-1} + i_t \odot \tilde{C}_t, \\
o_t &= \sigma\!\big(\Phi_o(v_t;\Theta_o)\big), \\
h_t &= o_t \odot \tanh(C_t).
\end{align}
\end{subequations}

\noindent\textbf{Remarks.}
\begin{itemize}
    \item Eq.~\eqref{eq:qkan_gate_map} follows the Kolmogorov--Arnold formulation,
    where node outputs are additive compositions of quantum nonlinear edge functions.
    \item Eq.~\eqref{eq:qvaf_output}--\eqref{eq:qvaf_circuit} define the QVAF,
    implemented through data re-uploading circuits whose variational parameters
    \(\boldsymbol{\theta}\) are trained end-to-end.
    \item Classical nonlinearities \(\sigma(\cdot)\) and \(\tanh(\cdot)\)
    are preserved for gating stability,
    while quantum activations enrich the inner functional space of each gate.
    \item The full trainable parameter set is
    \(\Theta_g = \{ \boldsymbol{\theta}_{g,p},\, a^{(\ell)},\, b^{(\ell)}\}\),
    jointly optimized with other network weights.
\end{itemize}

\subsubsection{Training and Optimization}

The training of QKAN-LSTM
model involves optimizing both classical and quantum parameters that define the nonlinear
mapping of each LSTM gate.

\paragraph{Loss Function}

For the regression task, the training objective is
to minimize the Mean Squared Error (MSE) between the predicted
and target outputs:
\begin{equation}
\mathcal{L} = \frac{1}{T}\sum_{t=1}^{T} (y_t - \hat{y}_t)^2,
\end{equation}
where \(y_t\) and \(\hat{y}_t\) denote the true and predicted values at time step \(t\).

\paragraph{Gradient Computation}

The model parameters
\(\Theta_g = \{\alpha_{g,p},\, a^{(\ell)},\, b^{(\ell)},\, \boldsymbol{\theta}_{g,p}\}\)
are optimized through hybrid quantum--classical backpropagation.
Gradients of the classical parameters \(\alpha_{g,p}\)
are computed using standard backpropagation through time (BPTT),
while the gradients of quantum parameters
\(\boldsymbol{\theta}_{g,p}\) are obtained via the parameter-shift rule
\cite{wierichs2022general}:
\begin{equation}
\label{eq:param_shift_split}
\begin{split}
\frac{\partial \phi_{g,p}(u;\boldsymbol{\theta})}{\partial \theta_k}
= \frac{1}{2}
\Big[
&\phi_{g,p}\!\left(u;\boldsymbol{\theta}_k+\tfrac{\pi}{2}\mathbf{e}_k\right) \\
\quad &-\,
\phi_{g,p}\!\left(u;\boldsymbol{\theta}_k-\tfrac{\pi}{2}\mathbf{e}_k\right)
\Big].
\end{split}
\end{equation}
where $\mathbf{e}_k$ denotes the unit vector indicating the $k$-th parameter
in $\boldsymbol{\theta}$, ensuring that only $\theta_k$ is shifted while all
other parameters remain fixed.

In our implementation, the QKAN modules operate in the \emph{exact solver}
mode, where all QVAF are represented by analytic,
differentiable functions; therefore, gradients are computed directly using PyTorch’s autograd\cite{paszke2019pytorchimperativestylehighperformance}.

However, when executed on real quantum hardware or simulated quantum backends, the parameter-shift rule Eq.~\eqref{eq:param_shift_split} is employed to estimate the derivatives of quantum observables.

\paragraph{Optimization Algorithm}

A optimizer such as Adam\cite{kingma2014adam} or RMSprop\cite{kurbiel2017training}
is employed to jointly update both classical and quantum parameters.
At each iteration, the optimizer computes the gradient of the loss
with respect to all elements of \(\Theta_g\),
then updates them according to
\(\Theta_g \leftarrow \Theta_g - \eta \nabla_{\Theta_g}\mathcal{L}\),
where \(\eta\) is the learning rate.
This hybrid optimization loop iteratively refines both the quantum
variational circuits and classical combination weights until convergence.

\subsection{Jiang-Huang-Chen-Goan Networks}

\begin{figure}[th!]
\centering
\includegraphics[width=\columnwidth, trim={3cm 1cm 3cm 5cm}, clip]{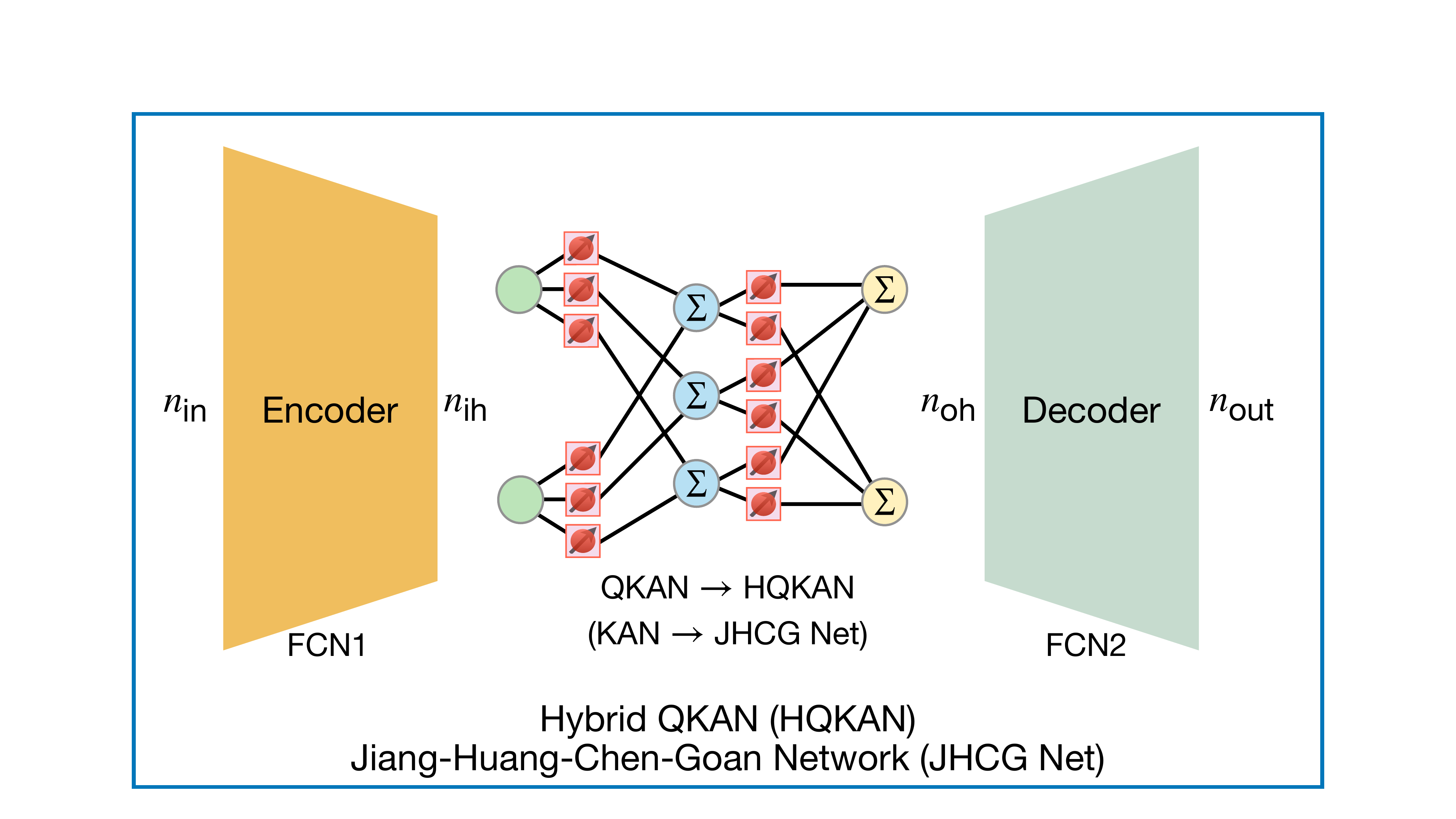}
\caption{
\textbf{Architecture of the Jiang–Huang–Chen–Goan Network (JHCG Net)~\cite{jiang2025qkan}.}
The JHCG Net comprises a fully connected encoder and decoder with a Kolmogorov–Arnold Network (KAN) serving as the latent feature processor, forming an autoencoder-like architecture.
When the latent KAN module is implemented using Quantum Kolmogorov–Arnold Networks (QKANs), the framework is referred to as the \textit{Hybrid QKAN (HQKAN)}, integrating quantum-inspired nonlinear transformations within the latent representation space.
}
\label{fig:hqkan}
\end{figure}

Ref.~\cite{jiang2025qkan} further introduced the concept of HQKANs, representing a fusion of classical and quantum-inspired neural computation. 
Building upon this foundation, the JHCG Net generalizes the KAN paradigm into a scalable, autoencoder-like framework designed for hierarchical representation learning. 
The architecture comprises three principal modules: a fully connected encoder, a latent KAN processor, and a decoder. 
The encoder compresses high-dimensional input features into a compact latent space, where the KAN module performs nonlinear transformations through parameterized univariate functions. 
The decoder subsequently reconstructs the output from the processed latent representation. 
By integrating the functional decomposition capability of KANs with the hierarchical abstraction mechanisms of deep neural networks, the JHCG Net achieves interpretable and parameter-efficient feature compression, transformation, and reconstruction.

Replacing the latent KAN processor with a QKAN yields the HQKAN architecture, which inherits quantum-inspired expressivity and frequency adaptability from the DARUAN activation framework while retaining the differentiability and scalability of classical optimization pipelines. 
The structural composition of HQKAN is illustrated in Figure~\ref{fig:hqkan}, emphasizing its encoder–KAN–decoder topology and the inclusion of a QKAN module within the latent layer. 
Unlike conventional QKANs that operate primarily as direct function approximators, HQKANs act as compositional operators embedded within classical architectures, integrating quantum-inspired nonlinear transformations into the latent feature space. 
Thus, this design enables an exponential enhancement in spectral capacity without increasing network width or depth, and facilitates high-fidelity, cross-modal representation learning with improved computational efficiency.

Crucially, HQKANs function as a \textit{scalable, drop-in substitute for MLPs} within modern deep architectures such as Transformers and Diffusion Models~\cite{jiang2025qkan}. 
They can replace conventional feed-forward layers in large-scale generative frameworks, providing superior expressivity, reduced parameterization, and enhanced convergence stability— while maintaining compatibility with classical GPU-based training pipelines.

Overall, the JHCG Net establishes a cohesive bridge between classical and quantum-inspired learning paradigms. 
While KANs offer interpretable and smooth functional decomposition, QKANs introduce exponentially compact yet expressive nonlinear mappings. 
The resulting HQKAN architecture thereby defines a unified, interpretable, and hardware-efficient foundation for scalable hybrid learning systems.

\section{Results}

\begin{table*}[t]
\centering
\caption{Number of parameters in LSTM, QLSTM, QKAN-LSTM and HQKAN-LSTM models.}
\label{tab:num_parameter}
\resizebox{\textwidth}{!}{
\begin{tabular}{|c|c|c|c|c|c|c|c|c|c|c|c|c|}
\hline
 & \multicolumn{3}{c|}{\textbf{LSTM}} &
   \multicolumn{3}{c|}{\textbf{QLSTM}} &
   \multicolumn{3}{c|}{\textbf{QKAN-LSTM}} &
   \multicolumn{3}{c|}{\textbf{HQKAN-LSTM}}\\
\hline

\textbf{Dataset} & \textbf{Classical} & \textbf{Quantum} & \textbf{Total} &
\textbf{Classical} & \textbf{Quantum} &\textbf{Total} &
\textbf{Classical} & \textbf{Quantum} &\textbf{Total} &
\textbf{Classical} & \textbf{Quantum} &\textbf{Total} \\
\hline
Damped SHM & 166 & - & 166 & 6 & 72 & 78 & 21 & 96 & 117 & 12 & 28& 40\\
\hline
Bessel Function & 166 & - & 166 & 6 & 72 & 78& 26 & 32 & 58& 25 & 8& 33\\
\hline
Urban Telecommunication & 277  & - & 277 & 5 & 100 & 105& 26 & 32 & 58& 36 & 53 & 89\\
\hline
\end{tabular}
}
\end{table*}

\begin{table}[h]
    \caption{Performance comparison on damped SHM dataset.}
    \label{tab:shm}
    \centering
    \begin{tabular}{@{}ccccc@{}}
        \toprule
        \textbf{Model} & 
        \textbf{Epoch} &
        \textbf{Training Loss} & 
        \textbf{Testing Loss} & 
        \textbf{\( R^2 \)} \\
        \midrule
        \multirow{3}{*}{LSTM} 
        & 1  & \(1.22 \times 10^{-1}\) & \(9.03 \times 10^{-3}\) & 0.7973 \\
        & 15 & \(5.86 \times 10^{-3}\) & \(1.38 \times 10^{-3}\) & 0.9690 \\
        & 30 & \(4.61 \times 10^{-3}\) & \(1.33 \times 10^{-3}\) & 0.9701 \\
        \midrule
        \multirow{3}{*}{QLSTM} 
        & 1  & \(1.59 \times 10^{-1}\) & \(1.84 \times 10^{-2}\) & 0.5855 \\
        & 15 & \(1.19 \times 10^{-2}\) & \(1.42 \times 10^{-3}\) & 0.9680 \\
        & 30 & \(1.74 \times 10^{-3}\) & \(1.24 \times 10^{-4}\) & 0.9972 \\
        \midrule
        \multirow{3}{*}{QKAN-LSTM} 
        & 1  & \(2.19 \times 10^{-1}\) & \(4.50 \times 10^{-2}\) & -0.0111 \\
        & 15 & \(3.20 \times 10^{-3}\) & \(1.92 \times 10^{-3}\) & 0.9569 \\
        & 30 & \(3.09 \times 10^{-3}\) & \(1.02 \times 10^{-3}\) & 0.9771 \\
        \midrule
        \multirow{3}{*}{HQKAN-LSTM} 
        & 1  & \(2.26 \times 10^{-1}\) & \(4.64 \times 10^{-2}\) & -0.0418 \\
        & 15 & \(1.65 \times 10^{-2}\) & \(2.72 \times 10^{-3}\) & 0.9391 \\
        & 30 & \(5.94 \times 10^{-3}\) & \(4.32 \times 10^{-4}\) & 0.9903 \\
        \bottomrule
    \end{tabular}
\end{table}

\begin{table}[h]
    \caption{Performance comparison on Besssel function dataset.}
    \label{tab:Bessel}
    \centering
    \begin{tabular}{@{}ccccc@{}}
        \toprule
        \textbf{Model} & 
        \textbf{Epoch} &
        \textbf{Training Loss} & 
        \textbf{Testing Loss} & 
        \textbf{\( R^2 \)} \\
        \midrule
        \multirow{3}{*}{LSTM} 
        & 1  & \(5.18\times10^{-2}\) & \(2.07\times10^{-3}\) & 0.9115 \\
        & 15 & \(2.47\times10^{-3}\) & \(9.38\times10^{-4}\) & 0.9471 \\
        & 30 & \(1.80\times10^{-3}\) & \(7.69\times10^{-4}\) & 0.9673 \\
        \midrule
        \multirow{3}{*}{QLSTM} 
        & 1  & \(8.32\times10^{-2}\) & \(2.53\times10^{-3}\) & 0.8923 \\
        & 15 & \(5.28\times10^{-3}\) & \(1.02\times10^{-3}\) & 0.9566 \\
        & 30 & \(3.40\times10^{-3}\) & \(7.53\times10^{-4}\) & 0.9679 \\
        \midrule
        \multirow{3}{*}{QKAN-LSTM} 
        & 1  & \(1.35\times10^{-1}\) & \(1.43\times10^{-2}\) & 0.3872 \\
        & 15 & \(2.99\times10^{-3}\) & \(6.12\times10^{-4}\) & 0.9739 \\
        & 30 & \(1.67\times10^{-3}\) & \(3.27\times10^{-4}\) & 0.9861\\
        \midrule
        \multirow{3}{*}{HQKAN-LSTM} 
        & 1  & \(1.09\times10^{-1}\) & \(6.61\times10^{-3}\) & 0.7185 \\
        & 15 & \(2.26\times10^{-3}\) & \(4.17\times10^{-4}\) & 0.9822 \\
        & 30 & \(1.86\times10^{-3}\) & \(3.21\times10^{-4}\) & 0.9863\\
        \bottomrule
    \end{tabular}
\end{table}

\begin{table*}[t]
\centering
\caption{
Comparison of MAE / MSE for different models across sequence lengths on Urban Telecommunication Dataset.\\
{\upshape
The best result is highlighted in \textbf{bold} while the second is labeled with \underline{underline}.
}
}
\label{tab:MSE_MAE}
\begin{tabular}{c|cccccc}
\toprule
Model & Seq\_len 4 &  Seq\_len 8 & Seq\_len 12 & Seq\_len 16 & Seq\_len 32 & Seq\_len 64 \\
\midrule
LSTM 
& 1.0633 / 4.7135 
& 1.0757 / 4.7011 
& 1.0799 / 4.6085 
& 1.0914 / 4.7020 
& 1.1211 / 4.8381 
& 1.1597 / 4.8853 \\

QLSTM 
& 1.0322 / \underline{4.5217}
& \underline{1.0324} / \textbf{4.5307} 
& 1.0466 / 4.5715 
& 1.0456 / 4.6244 
& 1.0634 / 4.5953 
& \underline{1.0933} / 4.7194 \\

\underline{QKAN-LSTM}
& \underline{1.0292} / \textbf{4.4377} 
& 1.0399 / \underline{4.5441} 
& \underline{1.0443} / \underline{4.5570}
& \underline{1.0418} / \textbf{4.5485} 
& \underline{1.0534} / \underline{4.5647} 
& 1.1103 / \underline{4.7311} \\

\textbf{HQKAN-LSTM} 
& \textbf{1.0045} / 4.5471
& \textbf{1.0249} / 4.6166
& \textbf{1.0361} / \textbf{4.5241} 
& \textbf{1.0189} / \underline{4.5985}
& \textbf{1.0378} / \textbf{4.4970}
& \textbf{1.0848} / \textbf{4.6749} \\
\bottomrule
\end{tabular}
\end{table*}

\subsection{Training Data}
We evaluate the proposed QKAN-LSTM and HQKAN-LSTM model on three representative datasets:
Damped Simple Harmonic Motion, Bessel Function, and Urban Telecommunication time-series data for real-world testing.

\begin{figure}[h]
    \centering
    \includegraphics[width=0.48\textwidth]{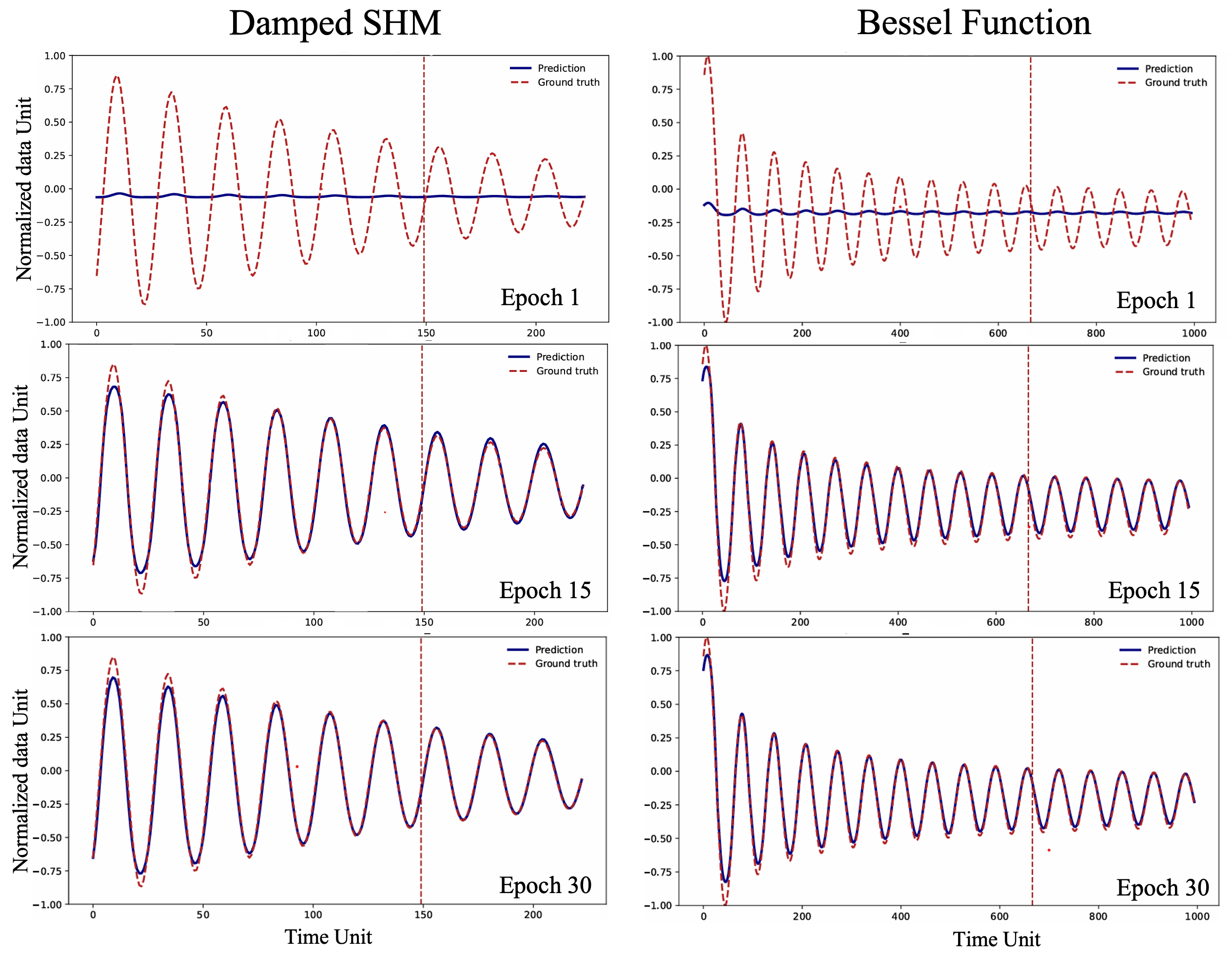}
\caption{Results of QKAN-LSTM on damped SHM and Bessel function datasets.}
    \label{fig:Result}
\end{figure}

\paragraph{Damped Simple Harmonic Motion}
The Damped Simple Harmonic Motion (Damped SHM) dataset represents a fundamental form of classical dynamics, 
describing oscillatory motion that exhibits an exponential decay in amplitude over time. 
It follows the canonical second-order differential equation 

\begin{equation}
\frac{d^2x(t)}{dt^2} + 2\zeta\omega_0\frac{dx(t)}{dt} + \omega_0^2 x(t) = 0 ,
\end{equation} 
where \( x(t) \) denotes the displacement at time \( t \), 
\( \omega = 2\pi f \) is the angular frequency of oscillation, 
and \( \zeta = \frac{c}{2mk} \) represents the damping ratio. 
We constructed a time-series dataset of damped SHM systems, 
where each sample includes the temporal variable \( t \) and the corresponding displacement \( x(t) \). 
This dataset serves as a benchmark for evaluating the model’s capability to learn and predict harmonic oscillatory dynamics.

\paragraph{Bessel Function}
The Bessel Function dataset represents a class of nonlinear oscillatory dynamics 
that frequently arises in the solutions of wave propagation and diffusion problems 
under cylindrical or spherical coordinate systems. 
It is governed by Bessel’s differential equation: 
\begin{equation}
x^2 \frac{d^2y(x)}{dx^2} + x \frac{dy(x)}{dx} + (x^2 - \alpha^2) y(x) = 0,
\end{equation}
where \( y(x) \) denotes the displacement (or amplitude) at position \( x \), 
and \( \alpha \) is the order of the Bessel function. 
The solution is given by the Bessel function of the first kind:
\begin{equation}
J_{\alpha}(x) = \sum_{m=0}^{\infty} \frac{(-1)^m}{m! \, \Gamma(m + \alpha + 1)} 
\left( \frac{x}{2} \right)^{2m + \alpha},
\end{equation}
where \( \Gamma(x) \) is the Gamma function. 
In this work, we constructed a time-series dataset based on the second-order Bessel function \( J_2(x) \), 
where each sample includes the variable \( x \) and its corresponding amplitude \( J_2(x) \). 

\paragraph{Urban Telecommunication}
The Urban Telecommunication dataset is derived from the Milan Telecommunication Activity Dataset~\cite{barlacchi2015multi}, 
which records telecommunication activities sampled every 10 minutes across a spatial grid of the city. 
To mitigate data sparsity and modality imbalance present in the original dataset, 
we focus exclusively on the univariate SMS-in channel, which provides higher completeness and reliability across grid cells. 
Follwing Ref.~\cite{chen2025benchmarking}, We preprocess the dataset by selecting grid cells that exhibit sufficient temporal continuity 
and normalize all SMS-in activity values to the interval \([0, 1]\). 
Each training instance is constructed from a fixed-length input sequence 
\(\mathbf{X} = [x_{t-T+1}, \ldots, x_t]\), 
with the subsequent value \(x_{t+1}\) used as the prediction target. 
To assess the model’s capability of capturing temporal dependencies across different horizons, 
we experiment with multiple sequence lengths \( T \in \{4, 8, 12, 16, 32, 64\} \). 
The dataset is divided into 70\% for training, 15\% for validation, and 15\% for testing.

\subsection{Experiment setup}
In our experiment, we evaluate the performance of LSTM, QLSTM, QKAN-LSTM, and HQKAN-LSTM models on three datasets: Damped SHM, Bessel Function, and Urban Telecommunication.
Experiments are simulated with PennyLane \cite{bergholm2022pennylaneautomaticdifferentiationhybrid}, PyTorch \cite{paszke2019pytorchimperativestylehighperformance}, and QKAN adapted from the open-sourced library on GitHub~\cite{jiang2025qkan_github}\footnote{Available at \href{https://github.com/Jim137/qkan}{https://github.com/Jim137/qkan}.}.
For implementations of LSTM and QLSTM, we followed the setup in ref.~\cite{chen2025benchmarking}.

For training, the learning rate is set to \(10^{-2}\) for the Damped SHM and Bessel Function datasets, and \(10^{-3}\) for the Urban Telecommunication dataset, except for HQKAN-LSTM, which uses a slightly higher rate of \(2\times10^{-3}\) to facilitate faster convergence.

For the LSTM model, the hidden unit size is set to 5 for the Damped SHM and Bessel Function datasets, and 4 for the Urban Telecommunication dataset. Similarly, for the QLSTM model, the hidden unit size is set to 5 for the Damped SHM and Bessel Function datasets, and 4 for the Urban Telecommunication dataset. In contrast, for the QKAN-LSTM and HQKAN-LSTM model, the hidden unit size is set to 1 for both the Bessel Function and Urban Telecommunication datasets, and 2 for the Damped SHM dataset. The input and output dimensions for all models are both set to 1, as these hyperparameter configurations are chosen to balance model expressivity and parameter efficiency, while ensuring a fair performance comparison across all architectures and properly reflecting their differences in representational capacity.

Regarding the quantum components, the QLSTM model utilizes 6 qubits for Damped SHM and Bessel Function, and 5 qubits for Urban Telecommunication. The quantum gates are parameterized with RY encoding, defined as
$RY(x) = e^{-i \frac{x}{2} \sigma_y}$, where $\sigma_y$ is the Pauli-Y operator and $x$ denotes the input data. Following the encoding stage, the circuit applies repeated CNOT + trainable $RY(\theta)$ blocks as the RealAmplitudes Anstaz~\cite{chen2025benchmarking,iwakiri2025universality} and performs Pauli-Z measurements to generate the quantum outputs. In contrast, both QKAN-LSTM and HQKAN-LSTM models employ a single-qubit DARUAN layer across all datasets.

The experiment on the Urban Telecommunication dataset, the models are trained for 50 epochs, with sequence lengths of \{4, 8, 12, 16, 32, 64\}, and our all experiments were carried out on a system equipped with NVIDIA Tesla V100 GPUs (16GB) and Intel Xeon CPUs.

\subsection{Result Analysis}

Table~\ref{tab:num_parameter} summarizes the number of classical and quantum parameters across all models. Compared with QLSTM, both QKAN-LSTM and HQKAN-LSTM achieve a notable reduction in quantum parameters on the Urban Telecommunication dataset---approximately 50--70\% fewer---while utilizing the single qubit DARUAN layer, yet maintaining comparable or superior predictive performance. In addition, compared with LSTM, both QKAN-LSTM and HQKAN-LSTM exhibit a substantial decrease in the total number of parameters on the same dataset. For the Damped SHM task, however, the relative simplicity of the sequence pattern limits the learning capacity of QKAN-LSTM when using a single hidden unit. Therefore, two hidden units are adopted in this case to ensure sufficient expressive power, which leads to a higher parameter count compared to the same model applied to other datasets. These results collectively highlight the parameter efficiency and scalability of the QKAN-based architecture, enabling effective sequence modeling with limited quantum resources.

The evaluation on the Damped SHM and Bessel Function datasets is presented in Table~\ref{tab:shm}, Table~\ref{tab:Bessel}, and Figure~\ref{fig:Result}, respectively. Across both datasets, all quantum-enhanced models exhibit steady convergence and high predictive accuracy as training progresses. 

In the Damped SHM task, QKAN-LSTM exhibits slower convergence during the early training stage, which can be attributed to the relatively simple oscillatory dynamics of the dataset. As training progresses, however, the model rapidly stabilizes and achieves superior accuracy, reflecting its adaptability even in low-complexity temporal patterns. As shown in Table~\ref{tab:shm}, QKAN-LSTM attains a final testing loss of \(1.02 \times 10^{-3}\) and an \(R^2\) score of 0.9771 after 30 epochs, surpassing LSTM baselines. Similarly, HQKAN-LSTM achieves comparable convergence with an \(R^2\) of 0.9903, confirming the hybrid model’s ability to sustain high predictive accuracy while maintaining a reduced number of quantum parameters.

For the Bessel Function dataset, QKAN-LSTM and HQKAN-LSTM demonstrate more robust and stable performance than LSTM and QLSTM, achieving testing losses of \(3.27 \times 10^{-4}\) and \(3.21 \times 10^{-4}\), respectively, with corresponding \(R^2\) scores exceeding 0.986, as shown in Table~\ref{tab:Bessel}. 

For the Urban Telecommunication dataset, Table~\ref{tab:MSE_MAE} presents the comparison of mean absolute error (MAE) and MSE values across different sequence lengths.
Overall, QKAN-LSTM and HQKAN-LSTM exhibit consistently lower error metrics than LSTM and QLSTM, demonstrating superior stability and adaptability across varying temporal dependencies.
Notably, from Table~\ref{tab:num_parameter}, both QKAN-based models employ substantially fewer quantum parameters than QLSTM and significantly fewer classical parameters than LSTM, highlighting their efficiency in balancing representational capacity with computational economy.
As the sequence length increases, both models maintain clear performance advantages, indicating scalability and the ability to capture long-range temporal correlations without significant degradation in accuracy. 
In particular, HQKAN-LSTM achieves the lowest MAE from short to long sequence lengths across the experiments, confirming the robustness of the hybrid quantum–classical design.
Collectively, these results validate that the QKAN-based architectures not only enhance predictive precision but also achieve efficient parameter utilization, demonstrating their practicality for complex real-world temporal datasets.

\section{Discussion}
In this work, the performance gains over the classical baseline are consistently associated with replacing the gate-wise affine maps in LSTM by DARUAN-based QVAFs.
DARUAN implements a single-qubit data re-uploading circuit with trainable pre-processing weights, yielding a compact and highly expressive univariate nonlinearity whose controllable Fourier spectrum can expand exponentially with the number of re-uploading repetitions~\cite{jiang2025qkan}.
When deployed inside recurrent gates, this spectral adaptability is particularly beneficial for time series exhibiting mixed periodicity, burstiness, and multi-scale temporal patterns, as the gate transformations can represent richer nonlinear filtering behavior without increasing hidden width.

Beyond expressivity, integrating QKAN into LSTM provides a structural advantage inherited from KAN-style networks: each gate transformation is realized as an additive composition of learnable univariate functions.
This decomposition offers a more transparent mechanism for analyzing how individual input channels and hidden-state components influence the forget/input/output decisions, enabling gate-level interpretability and more targeted diagnostics of the learned temporal control policy.
Compared to monolithic dense transformations, such edge-wise QVAF parameterization also encourages parameter reuse across time steps and alleviates redundancy in gate modeling.

From an optimization perspective, QKAN activations are computed as bounded expectation values of single-qubit observables, which naturally constrains gate pre-activations and acts as an implicit regularizer.
In recurrent learning, where backpropagation through time is sensitive to unstable gate dynamics, these bounded nonlinearities complement the outer sigmoid/tanh nonlinearities to improve training stability.
Moreover, because QKAN avoids multi-qubit entanglement by construction, it reduces dependence on deep entangling circuit structures and the associated hardware and optimization burdens common in VQC-based recurrent models~\cite{chen2022quantum}.
In particular, scaling multi-qubit VQCs often amplifies barren plateau phenomena~\cite{larocca2025barren} and exacerbates the impact of imperfect two-qubit gates~\cite{smith2025singleQubit, singh2024virtualTwoq} and shot-noise-dominated measurements~\cite{mitarai2018quantumBP,preskill2018NISQera,koch2020NISQchallenge,scriva2024challenges,cerezo2022challenges}.
By contrast, the single-qubit and additive structure of QKAN mitigates these issues and supports progressive expressivity scaling via the layer extension strategy (increasing the re-uploading depth to enrich frequency content while preserving previously learned features)~\cite{jiang2025qkan}, which is well aligned with stable optimization in recurrent settings.

Finally, QKAN-based gates can improve practical efficiency.
Since each DARUAN operates on a two-level quantum state, the resulting computations are lightweight and can be implemented with GPU-friendly tensor operations, enabling efficient classical simulation and deployment while maintaining quantum-inspired expressivity~\cite{jiang2025qkan}.
In our experiments, this translates into a favorable accuracy--efficiency trade-off: QKAN-LSTM and its autoencoder-like HQKAN-LSTM variant achieve competitive or superior forecasting performance while substantially reducing the number of trainable parameters relative to standard LSTM and requiring fewer quantum resources than multi-qubit QLSTM baselines.
Furthermore, ref.~\cite{jiang2025qkan} shows that HQKAN can serve as a drop-in replacement for MLP layers with a substantially reduced parameter count and lower training memory, while maintaining comparable training throughput.
In our setting, the HQKAN-LSTM inherits this efficiency advantage by compressing the latent dimensionality of the gate transformations, thereby reducing the overall parameterization and the corresponding memory footprint relative to a standard LSTM.

\section{Conclusion}
This study has demonstrated that incorporating the QKAN framework into LSTM architectures substantially enhances temporal sequence modeling capabilities across both synthetic and real-world datasets. 
The proposed QKAN-LSTM and HQKAN-LSTM models utilize the expressive power of quantum mappings to enrich classical recurrent representations, creating a compact yet powerful feature encoding for temporal dependencies. This is achieved by enhancing frequency adaptability and enabling an exponentially enriched spectral representation without requiring complex multi-qubit entanglement. This integration allows the models to deliver higher predictive accuracy and faster convergence while requiring significantly fewer trainable parameters and quantum resources, fundamentally boosting computational efficiency and scalability.

Looking ahead, the proposed QKAN-based LSTM models hold great potential to bridge the gap between classical and quantum computing paradigms, enabling efficient deployment in edge computing environments and resource-limited quantum devices. This advancement highlights the versatility and scalability of QKAN-enhanced architectures, positioning them as a key solution for both large-scale, complex, and high-dimensional datasets and low-resource quantum machine learning implementations in real-world settings.

\ifarxiv
\section*{Acknowledgment}
Y.C. Hsu, J.-C. Jiang, C.-H. Lin and K.C. Peng thank the National Center for High-Performance Computing (NCHC), National Institutes of Applied Research (NIAR), Taiwan, for providing computational and storage resources supported by National Science and Technology Council (NSTC), Taiwan, under Grants No. NSTC 114-2119-M-007-013.
H.-S. Goan acknowledges support from the NSTC, Taiwan, under Grants No. NSTC 113-2112-M-002-022-MY3, No. NSTC 113-2119-M-002-021, No. NSTC 114-2119-M-002-018, No. NSTC 114-2119-M-002-017-MY3, and from the National Taiwan University under Grants No. NTU-CC-115L8937, No. NTU-CC115L893704 and No. NTU-CC-115L8512. H.-S. Goan is also grateful for the support of the “Center for Advanced Computing and Imaging in Biomedicine (NTU-115L900702)” through the Featured Areas Research Center Program within the framework of the Higher Education Sprout Project by the Ministry of Education (MOE), Taiwan, the support of Taiwan Semiconductor Research Institute (TSRI) through the Joint Developed Project (JDP) and the support from the Physics Division, National Center for Theoretical Sciences, Taiwan.
\fi

\bibliographystyle{ieeetr}
\bibliography{reference}

\end{document}